\documentstyle[pra,aps,preprint]{revtex}
\tightenlines 
\title{Strong nonlinear regime\\
of resonant four-wave mixing in a gas}
\author{S.A.Babin, E.V.Podivilov, D.A.Shapiro}
\address{Institute of Automation \& Electrometry,\\
Russian Academy of Sciences, Siberian Branch,\\
Novosibirsk, 630090, Russia}
\date{\today}
\begin{document}
\draft
\maketitle
\begin{abstract}
The explicit solution is obtained for four-wave mixing
$\omega_4=\omega_1-\omega_2+\omega_3$ of
two strong fields $\vec E_1, \vec E_3$
and two weak fields $\vec E_2, \vec E_4$ in a four-level system with
the large Doppler broadening.  The resonance of
the mixing coefficient dependence on intensity is found around
$\vec E_1\vec d_1=\vec E_3\vec d_3$, where $\vec d_{1,3}$
are the dipole moments of corresponding transitions. The effect
is interpreted as an intersection  of quasi-energy levels.
Up to 6 peaks appear in the dependence of conversion coefficient on
the detuning of the probe field $\vec E_2$. An unexpected additional
pair of peaks is a consequence of averaging over velocities and
disappears at low temperature. The results allow us to interpret
saturation behavior in recent experiments on the mixing in sodium
vapor.


\end{abstract}

Four-level system is a promising object for
resonant optics and spectroscopy owing to a great variety of
nonlinear effects. They are nonlinear interference, inversionless
gain, resonance refraction, electromagnetically induced
transparency, optically induced energy-level mixing and
shifting, population redistribution etc (see \cite{S92,K92}
and citations therein).  Recent experiments on continuous
four-wave frequency mixing of Raman type
with sodium molecules in a heat pipe
\cite{BHTW96,ABHTW97} gave interesting dependencies of generated wave
power on frequencies and intensities of the incident waves.
In particular, the dependence of output power on the first
strong field intensity was saturated in experiment
\cite{BHTW96} in thin media, whenever that on the third
wave intensity demonstrated the linear growth.   The
measurements were taken at large Doppler broadening while
the nonperturbative theory is proposed \cite{P96,CKHG96}
for atoms at rest.
The saturation of four-wave mixing efficiency as a function
of intensity was observed in optically thick lead vapor
interacted with megawatt pulses \cite{JXYMH96}. In case
the Rabi splitting exceeded the inhomogeneous width.

Development of nonperturbative theory from the mathematical
point of view involves the solution of the set of 16
algebraic equations for steady-state elements of atomic
density matrix for four-level system.
The problem is only of analyzing the resultant awkward
expression and to average it with Maxwellian distribution
over velocities. In present paper we study the particular
case of alternate two strong and two weak fields
interacting with 4-level system having some symmetry.  The
4th degree equation can be reduced to biquadratic
one, then the integration is possible analytically.

A simple explicit formula for nonlinear susceptibility
at zero frequency detuning displays the resonant
behavior as a function of a strong field at fixed another
strong field. We interpreted it as the coherent effect owing to
the intersection of quasi-energy levels. The susceptibility
as a function of weak field frequency
$\omega_2$, has 6, 4 or 3 peaks. The profile also displays the
important role of the Rabi splitting.

Let us consider the conversion of two strong incident waves
$\vec E_{1,3}$ resonantly interacting with opposite
transitions $gl, mn$ and the weak field $\vec E_2$ near
the resonance with transition $gn$ into the 4-th output
wave $\vec E_4$, inset of Fig.\ 1.  The electric field
in the cell is
\begin{equation} \vec E(\vec
r,t)=\sum_{\nu=1}^4 \vec E_\nu \exp\left(i\omega_\nu
t-i\vec k_\nu\vec r\right),
\label{field}
\end{equation}
where $\vec E_\nu$ is the amplitude of $\nu$-th field,
$\omega_\nu, \vec k_\nu$ are the frequency and wavevector.
The strong fields are also near resonance,
$\omega_1\simeq\omega_{gl},\omega_3\simeq\omega_{mn}$,
where $\omega_{ij}=\left(E_i-E_j\right)/\hbar$ are
transition frequencies between energy levels $E_i$ and
$E_j$. The intensity of the 4-th wave, that appears during
the process of mixing, is also being small. Its frequency
and wavevector satisfy the phase matching condition
\begin{equation}
\omega_4=\omega_1-\omega_2+\omega_3,\ \vec k_4=\vec
k_1-\vec k_2+\vec k_4.
\label{parametric}
\end{equation}

Maxwell equation for the output wave can be reduced to
\begin{equation}
{d\vec E_4\over dx}=-{2\pi i \omega_{ml}\vec d_{ml}\over c}
\left\langle \rho_{ml}\right\rangle,
\label{output}
\end{equation}
where $x$ is the coordinate, $\vec d_{ml}$ is the matrix
element of the dipole moment operator $\hat{\vec d}$, $c$
is the speed of light, $\rho_{ml}$ is the coherence at
transition $ml$, angular brackets denote the averaging
over velocity distribution. We should calculate $\rho_{ml}$
as a function of input amplitudes $\vec E_{1,2,3}$, their
wavevectors $\vec k_{1,2,3}$ and frequency detuning of the
weak field.

For this end we solve the equation for Wigner's atomic
density matrix (see \cite{RS91})
\begin{equation}
\left({\partial\over\partial t}+\vec
v\nabla+\gamma_{ij}\right) \rho_{ij}=q_j\delta_{ij}-i[\hat
V,\hat \rho]_{ij},
\label{Wigner}
\end{equation}
where $\vec v$ is the atomic velocity, $\gamma_{ij}$ are
relaxation constants, $q_j=Q_j\exp(-\vec
v^2/v_T^2)/v_T^3\pi^{3/2}$ is the Maxwellian excitation
function, $\hat V=-\vec E(\vec r,t)\hat{\vec d}/2\hbar$ is
the operator of interaction, $i,j=m,n,g,l$.

To the zeroth approximation we can neglect both the weak
fields $\vec E_{2,4}\to 0$. The set boils down to finding
out populations $\rho_j\equiv\rho_{jj}$ and coherences
$\rho_1\equiv\rho_{gl}\exp(-i\omega_1t+i\vec k_1\vec r)$,
$\rho_3\equiv\rho_{mn}\exp(-i\omega_3t+i\vec k_3\vec r)$
of a pair of separated two-level systems. The solution is
written as
\begin{eqnarray}
\rho_l&=&N_l-N_1{2|G_1|^2\gamma_1\over
\gamma_l\left(\Gamma_{s1}^2+\Omega_1^{\prime 2}\right)},
\nonumber\\
\rho_g&=&N_g+N_1{2|G_1|^2\gamma_1\over
\gamma_g\left(\Gamma_{s1}^2+\Omega_1^{\prime 2}\right)},
\label{two-level_1}\\
\rho_1&=&{iG_1N_1\Gamma_1^*\over
\Gamma_{s1}^2+\Omega_1^{\prime 2}};\nonumber
\end{eqnarray}
\begin{eqnarray}
\rho_n&=&N_n-N_3{2|G_3|^2\gamma_3\over
\gamma_n\left(\Gamma_{s3}^2+\Omega_3^{\prime 2}\right)},
\nonumber\\
\rho_m&=&N_m+N_3{2|G_3|^2\gamma_3\over
\gamma_m\left(\Gamma_{s3}^2+\Omega_3^{\prime 2}\right)},
\label{two-level_3}\\
\rho_3&=&{iG_3N_3\Gamma_3^*\over
\Gamma_{s3}^2+\Omega_3^{\prime 2}},\nonumber
\end{eqnarray}
where $N_j=q_j/\gamma_{jj}, j=m,n,l,g$ are the unperturbed
populations, $N_{1}=N_l-N_g$, $N_3=N_n-N_m$ are the
population differences at ``strong'' transitions,
$\gamma_1\equiv \gamma_{gl},\gamma_3=\gamma_{mn}$ are their
homogeneous width, $G_1=\vec E_1\vec d_{gl}/2\hbar$,
$G_3=\vec E_3\vec d_{mn}/2\hbar$ are the Rabi frequencies,
$\Omega'_\nu=\Omega_\nu-\vec k_\nu\vec v$ is the
Doppler-shifted detuning $\Omega_1=\omega_1-\omega_{gl}$,
$\Omega_3=\omega_3-\omega_{mn}$,
$\Gamma_\nu=\gamma_\nu+i\Omega'_\nu,$
\begin{eqnarray}
\Gamma_{s1}^2&=&\gamma_1^2+2|G_1|^2
\gamma_1(\gamma_g^{-1}+\gamma_l^{-1}),\nonumber\\
\Gamma_{s3}^2&=&\gamma_3^2+2|G_3|^2
\gamma_3(\gamma_m^{-1}+\gamma_n^{-1})\nonumber
\end{eqnarray}
are the homogeneous widths including the power
broadening.

Weak fields with amplitudes $G_2=\vec
E_2\vec d_{gn}/2\hbar, G_4=\vec E_4\vec d_{ml}/2\hbar$ lead
to appearance of cross-coherence between levels belonging
to the opposite two-level systems $\rho_2\equiv\rho_{gn},
\rho_4\equiv\rho_{ml}$ at the allowed transitions, as well
as at the forbidden transitions $\rho_5\equiv\rho_{gm},
\rho_6\equiv\rho_{nl}$. To the first order
one can neglect the influence of these fields to the
populations. The set of 4 algebraic equations appears for
the nondiagonal matrix elements:
\begin{eqnarray}
\Gamma_2 \rho_2 - iG_1 \rho_6^* + iG_3 \rho_5 &=&
-iG_2(\rho_g-\rho_n),\nonumber\\
\Gamma_4^* \rho_4^* + iG_3^* \rho_6^* - iG_1^* \rho_5 &=&
iG_4^*(\rho_m-\rho_l),\label{coherence}\\
\Gamma_5 \rho_5 - iG_1 \rho_4^* + iG_3^* \rho_2  &=&
iG_2\rho_3^* - iG_4^*\rho_1,\nonumber\\
\Gamma_6^* \rho_6^* + iG_3 \rho_4^* - iG_1^* \rho_2 &=&
-iG_2\rho_1^* + iG_4^*\rho_3.\nonumber
\end{eqnarray}
Here $\gamma_2\equiv\gamma_{gn}, \gamma_4\equiv\gamma_{ml}$
are the constants of relaxation of the coherence at the
allowed transition, $\gamma_5\equiv\gamma_{gm},
\gamma_6\equiv\gamma_{nl}$ are the constants for forbidden
transitions, $\Omega'_5=\Omega'_1-\Omega'_4,
\Omega'_6=\Omega'_1-\Omega'_2$ are frequency detunings at
these transitions.

The solution of Eq.\ (\ref{coherence}) for the nondiagonal
element at output transition $ml$ can be presented as
\begin{equation}\label{rho4}
\rho_4^*=-i\beta_4 G_1^*  G_2 G_3^* -i\alpha_4 G_4^*.
\end{equation}
During the initial step of the mixing the generated field
is small, $|G_4|\ll|G_2|$, that enables one to neglect
the absorption $\alpha_4$ and to find the coefficient
$\beta_4$ only.  We found intensity of output wave within
the thin medium approximation by the integration of Eq.\
(\ref{output}) from $x=0$ to the length of cell $L$
\begin{eqnarray} I_4(L)= {2\pi^2\omega_{ml}L\over c^2
\hbar^3} \left|\left\langle\beta_4\right\rangle (\vec
d_{gl}\vec e_1) (\vec d_{gn}\vec
e_2)\right.\times\nonumber\\ \left.(\vec d_{mn}\vec
e_3)(\vec d_{ml}\vec e_4)
\right|^2 I_1I_2I_3,\label{I4}
\end{eqnarray}
where $\vec e_\nu$
is the polarization of $\nu$-th wave, $I_\nu$ is its intensity.
We find coefficient $\beta_4$ comparing
Eq.\ (\ref{coherence}) to solution of
the form (\ref{rho4}).
\begin{eqnarray}
\beta_4={1\over D}\left(
(\Gamma_5+\Gamma_6^*)(\rho_g-\rho_n) -
{\rho_1^*\over iG_1^*}(|G_1|^2-|G_3|^2
\right.\nonumber\\
\left.-\Gamma_2\Gamma_5)-
{\rho_3^*\over iG_3^*}(|G_3|^2-|G_1|^2-\Gamma_2\Gamma_6^*)
\right).\label{beta4}
\end{eqnarray}
Here elements $\rho_g, \rho_n, \rho_1, \rho_3$ are defined
by Eq.\ (\ref{two-level_1}), (\ref{two-level_3}), the
determinant of set (\ref{coherence}) is
\begin{eqnarray}
D=\Gamma_2\Gamma_5\Gamma_4^*\Gamma_6^*
+ (|G_1|^2-|G_3|^2)^2 \nonumber \\
+ {1\over
2}(|G_1|^2+|G_3|^2)(\Gamma_2+\Gamma_4^*)(\Gamma_5+\Gamma_6^*)
\nonumber\\
-{1\over
2}(|G_1|^2-|G_3|^2)(\Gamma_2-\Gamma_4^*)(\Gamma_5-\Gamma_6^*),
\label{Determinant}
\end{eqnarray}
the polynomial of 4th degree in velocity. The
averaging of coefficient $\beta_4$ over velocity  is
possible by the residues theory for the Doppler limit
$v_T\to\infty$.

To examine the intensity dependence of coefficient
$\beta_4$ let us consider the case of equal relaxation
constants of the levels $\gamma_j=\gamma, j=g,l,m,n$,
excitation of the lower level only
$Q_j/\gamma=N\delta_{jl}$, the resonant strong field
detunings $\Omega_1/k_1=\Omega_3/k_3\ll v_T$, and equal
wavenumbers of both weak fields $|k_2-k_4|\ll
(k_2k_4k_5k_6)^{1/4}$. The last condition is natural in
down-conversion scheme. In view of phase matching
condition (\ref{parametric}) it is felt that the
weak field detunings depend on single parameter $\Omega$:
$\Omega_2=k_2\Omega_1/k_1+\Omega,
\Omega_4=k_4\Omega_1/k_1-\Omega$. Within the assumptions
one can also see that $k_4=k_2, k_6=k_5$. If all the
wavevectors are parallel, then the expression for
$\left\langle\beta_4\right\rangle$ assumes a simple form
\begin{eqnarray} \left\langle\beta_4\right\rangle={N\over
\sqrt{\pi}v_T}e^{-\Omega_1^2/k_1^2v_T^2}
\int\limits_{-\infty}^\infty {C(x)\over D(x)}{dx\over
\Gamma_{s1}^2 +k_1^2x^2},
\label{beta_4}\\
C(x)=4|G_1|^2iz+(\gamma-ik_1x)\times\nonumber\\
\left[|G_1|^2-|G_3|^2-\left(\gamma-i(k_2x-\Omega)
(\gamma-i(k_5x-\Omega)\right)\right].\nonumber
\end{eqnarray}
Here $x=\vec k_2\vec v/k_2-\Omega_1/k_1,$
$z=\Omega-i\gamma$,
$\Gamma_{s1}^2=\gamma^2+4|G_1|^2$ is the saturated width.
Determinant $D(x)$ turns to be a function of $x^2$
\begin{eqnarray}
D(x)=\kappa^4x^4-2\kappa^2x^2\Delta_1+\Delta_2^2,\quad
\kappa=\sqrt{k_2 k_5}
\label{determ_x2}\\
\Delta_1=(\mu^2/2-1)z^2-|G_1|^2+|G_3|^2,\quad \mu=k_1/\kappa\ge 2,
\nonumber\\
\Delta_2^2=\left[z^2-\left(|G_1|-|G_3|\right)^2\right]
\left[z^2-\left(|G_1|+|G_3|\right)^2\right].\nonumber
\end{eqnarray}
The detuning dependence of $|\Delta_2|$ takes the minimal
values at
\begin{equation}
\Omega=\pm |G_1|\pm|G_3|.
\label{resonances}
\end{equation}
It is a consequence of the level splitting by the strong
driving field. Note that at $|G_1|=|G_3|$ two points of
minimum merge together. The reason is equal Rabi splitting
for each level.

The simple form of the determinant (\ref{determ_x2}) allows
calculating mixing coefficient (\ref{beta_4}) explicitly
\begin{eqnarray}
\left\langle\beta_4\right\rangle={\sqrt{\pi}\over\kappa v_T}
{Ne^{-\Omega_1^2/k_1^2v_T^2}\over\Gamma_{s1}^2
+\Gamma_{s1}R\mu+\Delta_2\mu^2}
\left[{\gamma+iz\mu^2\over R}\right.\nonumber\\
+\left.{4iz|G_1|^2+\gamma(z^2+|G_1|^2-|G_3|^2)\over\Delta_2}
\left({1\over R}+{\mu\over\Gamma_{s1}}\right)\right],
\label{explicit}
\end{eqnarray}
where $R=\sqrt{2(\Delta_2-\Delta_1)}$, $\Re R>0$.
The branch of two-valued function $\Delta_2$ should be chosen
according to the following rules
\begin{eqnarray}
\Re\Delta_2<   0\ \mbox{at}\  P_+<|\Omega|,    \quad
\Re\Delta_2\geq0\ \mbox{at}\  |\Omega|\le P_-,\nonumber\\
{\rm sign}(\Im\Delta_2)={\rm sign}\ \Omega
\ \mbox{at}\ P_-< |\Omega| \le P_+,\nonumber
\end{eqnarray}
where $P_\pm=\left||G_1|\pm|G_3|\right|.$

The mixing coefficient $|\left\langle\beta_4\right\rangle|^2$
calculated from Eq. (\ref{explicit}) is plotted in Fig.\ 1 (a)
as a function of detuning
$\Omega$.  The coefficient
has 4 peaks at points given by (\ref{resonances}).
At equal distances between
quasi-energy levels $|G_1|=|G_3|$ two central peaks
coalesce in the center $\Omega=0$, Fig.1 (c). Except of the zeros of
$\Delta_2$, zeros of $R(\Omega)$ may add two peaks near the center,
Fig.\ 1 (b). The additional central peaks are absent for
motionless atoms since only four transitions are possible
between two pairs of splitted quasi-energy sublevels. These
peaks are contrast at $G_1>G_3\gg \gamma$ and disappear at
$|G_1/G_3|<\mu/\sqrt{\mu^2-1}$.

The value $|\left\langle\beta_4\right\rangle|^2$ at the
exact resonance $\Omega_\nu=0, \nu=1,\dots,4$ is shown in
Fig.\ 2 as a function of $|G_1|^2$.  The sharp peak at
$|G_1|=|G_3|$ confirms the qualitative interpretation of
the effect as the intersection of quasi-energy levels.
Inset in Fig.\ 2 illustrates why the maximal conversion
occurs when the Rabi splitting in opposite two-level
systems are equal.  Here the cross-transition from the
upper sublevel of level $m$ to the upper sublevel
of level $n$ has the same frequency as the transition
between  their lower sublevels.  In this case only 3
resonances remain in the spectrum, Fig.\ 1 (c), with the
overpowering maximum in the center.
The resonance condition $|G_1|=|G_3|$ brings
the maximum conversion efficiency in the intensity
dependence.

The splitting effect is evident from experimental results
on resonant four wave mixing in Na$_2$ \cite{BHTW96,ABHTW97}. The main feature
is the saturation of output power as a function of one
strong field. The conditions of experiment \cite{BHTW96} are
generally satisfy  the above model:  (1) down-conversion
level scheme $\omega_4<\omega_1$ (see inset, Fig.\ 1) with
$k_1v_T=7.0\cdot 10^9$~s$^{-1}$, $k_2v_T=6.5\cdot
10^9$~s$^{-1}$, $k_3v_T=5.2\cdot 10^9$~s$^{-1}$,
$k_4v_T=5.7\cdot 10^9$~s$^{-1}$;
(2) all incident waves 1,2,3 are generated by external lasers;
(3) the region of interaction is short enough (nearly $1$~cm), the
model of thin  media can be treated;
(4) estimated level parameters are
$N_l\sim 10^{12}\mbox{\ cm}^{-3}\gg
N_n\sim 10^{11}\mbox{\ cm}^{-3} \gg N_g, N_m$,
$\gamma_m\simeq \gamma_g\sim 2\cdot 10^8\mbox{\ s}^{-1}$,
$\gamma_n\simeq \gamma_l\sim 2\cdot 10^7\mbox{\ s}^{-1}$.
Slightly noncollinear geometry (mixing angle $\theta\sim
10^{-2}$) leads to an effective broadening $\Delta\omega\sim
kv_T\cdot\theta\sim 10^8\mbox{\ s}^{-1}$. Another factor is usual
jitter of laser frequencies, especially for dimer and dye lasers,
$\Delta\omega\sim (2\div 4)\cdot 10^8\mbox{\ s}^{-1}$.
Thus, the effective value $\gamma=(3\div 6)\cdot
10^8\mbox{\ s}^{-1}$ seems reasonable; (5) the maximal
field values estimated from the focusing geometry
$|G_1|^{\max}\sim 10^9\mbox{\ s}^{-1}$,
$|G_2|^{\max}\sim 2\cdot 10^8\mbox{\ s}^{-1}$,
$|G_3|^{\max}\sim 5\cdot 10^8\mbox{\ s}^{-1}$ nearly
correspond to the condition of two strong fields.

The resonance condition $|G_1|=|G_3|$ may result in peaks
as in dependence $\beta_4(I_1)$, as in dependence
$\beta_4(I_3)$. If $|G_1|^{\max}>|G_3|^{\max}$, the peak is
seen only in $\beta_4(I_1)$. The width of the peak is
determined by the decay rate $\gamma$. Since in the
experiment $\gamma\sim|G_3|$, the peak is wide, Fig.\ 2(b),
and gives a smooth saturation curve $I_4(I_1)$, Fig.\ 2(c).
According to this consideration the saturation of
$I_4(I_1)$ in the experiment (boxes in Fig.\ 2) is observed
at $|G_1|>|G_3|$ and
there is no saturation for $I_4(I_3)$.  Note that such
behavior was observed for different values of $I_2$ varied
by one order.  Under the opposite experimental condition \cite{ABHTW97}
$|G_1|^{\max} <|G_3|^{\max}$ the dependencies $I_4(I_1)$
and $I_4(I_3)$ change their behavior in agreement with the
consideration.

Thus, the model explains quantitatively the main features
of the measured saturation curves. To observe the sharp
resonances arising from Rabi splitting the stabilization of
laser frequencies seems to be important. To increase the
efficiency of conversion into the 4th wave it is necessary
to tune up the laser frequencies to corresponding peaks.
The optimum at $\Omega_\nu=0$ corresponds to equal Rabi
frequencies $|G_1|=|G_3|$.

Authors are grateful to S.G.~Rautian, A.M.~Shalagin, and
M.G.~Stepanov for fruitful discussions, B.~Wellegehausen and
A.A.~Apolonsky for clarifying the details of experiments.
This work was partially supported by Deutsche
Forschungsgemeinschaft, grant WE 872/18-1.


\section*{List of Captions}

Fig. 1. Conversion coefficient
$|\left\langle\beta_4\right\rangle|^2$ (arb. units) as a
function of detuning $\Omega$ of the second field
at $|G_1|=1$, $|G_3|=0.5$, $k_1v_T=7.0$,
$k_2v_T=6.9$, $\gamma=0.2$
(a), $\gamma=0.02$ (b), and $\gamma=0.02$
at $|G_1|=|G_3|=0.5$ (c)
(all frequencies are
in $\mbox{ns}^{-1}$).
Inset is the level diagram of four-level
system interacting with two strong driving fields at the
opposite transitions (solid arrows) and two weak fields
(wavy arrows). Dotted lines show the forbidden transitions.

Fig. 2. Conversion coefficient
$|\left\langle\beta_4\right\rangle|^2$ (arb. units) vs
$|G_1|^2$ at $|G_3|=0.5$, $\Omega=0$, $k_1v_T=7$,
$k_2v_T=6.5$:  $\gamma=0.06$ (a), $\gamma=0.6$
(b), and $|G_4|^2$ vs $|G_1|^2$ at $\gamma=0.6$ (c). The
parameters for (b), (c) correspond to experiment, all frequencies are
in $\mbox{ns}^{-1}$.
Boxes denote the experimental points from \cite{BHTW96}.  The
inset illustrates the Rabi splitting of dressed states.

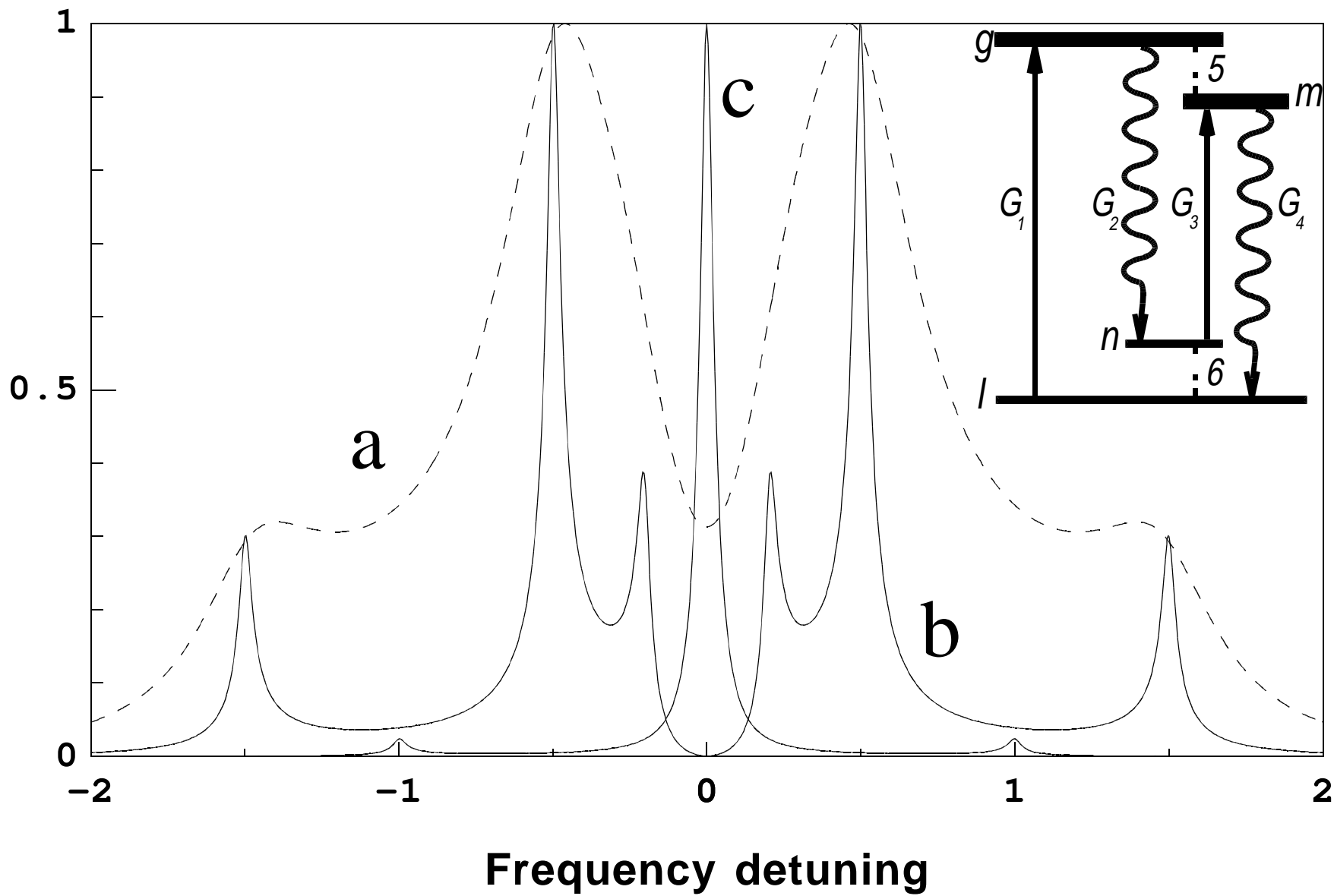
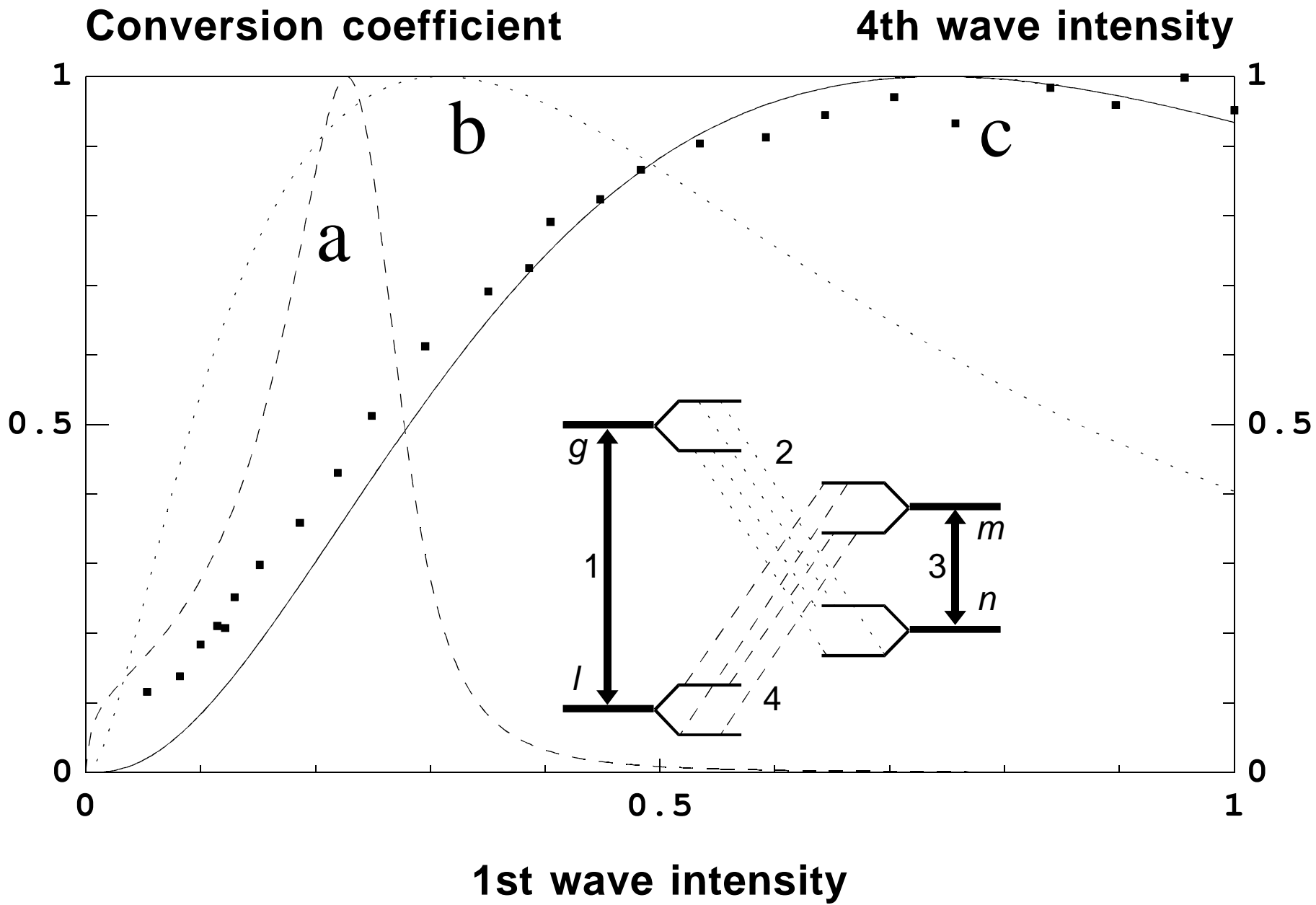
\end{document}